\documentclass[cits]{PoS}
\usepackage[authoryear,square]{natbib}
\bibpunct{(}{)}{;}{a}{}{,}
\newcommand{\apj}{ApJ}
\newcommand{\apjl}{ApJL}
\newcommand{\mnras}{MNRAS}
\newcommand{\nar}{New Astron. Rev.}
\newcommand{\nat}{Nature}
\newcommand{\prd}{Phys. Rev. D}
\newcommand{\PKU}{School of Physics, Peking University, Beijing
  100871, China} 
\newcommand{\UBC}{Department of Physics and Astronomy, University of
  British Columbia, 6224 Agricultural Rd., Vancouver, BC V6T 1Z1,
  Canada} 
\newcommand{\MPIfR}{Max-Planck-Institut f\"ur Radioastronomie, Auf dem
  H\"ugel 69, D-53121 Bonn, Germany}
\newcommand{\Swinburne}{Centre for Astrophysics and Supercomputing and 
  ARC Centre for All-Sky Astrophysics (CAASTRO), Swinburne University
  of Technology, PO Box 218 Hawthorn, VIC 3122, Australia}

\newcommand{\ASTRON}{ASTRON, The Netherlands Institute for Radio
  Astronomy, 7990 AA Dwingeloo, The Netherlands}
\newcommand{\APIA}{Anton Pannekoek Institute for Astronomy, University
  of Amsterdam, Science Park 904, 1098 XH Amsterdam, The Netherlands}
\newcommand{\Oldenburg}{University of Oldenburg, Department of
  Physics, 26111 Oldenburg, Germany} 
\newcommand{\Bremen}{University of Bremen, ZARM, 28359 Bremen,
  Germany} 
\newcommand{\OAC}{INAF-Osservatorio Astronomico di Cagliari, via della
  Scienza 5, 09047 Selargius (CA), Italy} 
\newcommand{\NRAO}{National Radio Astronomy Observatory, 520 Edgemont 
  Road, Charlottesville, VA 22903, USA}
\newcommand{\JBCA}{Jodrell Bank Centre for Astrophysics, The
  University of Manchester, M13 9PL, United Kingdom}  
\title{Testing Gravity with Pulsars in the SKA Era}
\ShortTitle{Testing gravity with pulsars in the SKA Era}
\author{
\speaker{Lijing Shao}$^1$,
{Ingrid H. Stairs}$^2$,
{John Antoniadis}$^3$,
{Adam T. Deller}$^4$,
{Paulo C. C. Freire}$^3$,
{Jason W. T. Hessels}$^{4,5}$,
{Gemma H. Janssen}$^4$,
{Michael Kramer}$^{3,6}$,
{Jutta Kunz}$^7$,
{Claus L\"{a}mmerzahl}$^8$,
{Volker Perlick}$^8$,
{Andrea Possenti}$^9$,
{Scott Ransom}$^{10}$,
{Benjamin W. Stappers}$^6$,
{Willem van Straten}$^{11}$
\\
$^1$\PKU{}\\
$^2$\UBC{}\\
$^3$\MPIfR{}\\
$^4$\ASTRON{}\\
$^5$\APIA{}\\
$^6$\JBCA{}\\
$^7$\Oldenburg{}\\
$^8$\Bremen{}\\
$^9$\OAC{}\\
$^{10}$\NRAO{}
$^{11}$\Swinburne{}\\
\\
E-mails: 
\email{lshao@pku.edu.cn} {\rm (LS);}
\email{stairs@astro.ubc.ca} {\rm (IHS)}
}
\abstract{ The Square Kilometre Array (SKA) will use pulsars to enable
  precise measurements of strong gravity effects in pulsar systems,
  which yield tests of gravitational theories that cannot be carried
  out anywhere else.  The Galactic census of pulsars will discover
  dozens of relativistic pulsar systems, possibly including pulsar --
  black hole binaries which can be used to test the ``cosmic
  censorship conjecture'' and the ``no-hair theorem''.  Also, the
  SKA's remarkable sensitivity will vastly improve the timing
  precision of millisecond pulsars, allowing probes of potential
  deviations from general relativity (GR). Aspects of gravitation to
  be explored include tests of strong equivalence principles,
  gravitational dipole radiation, extra field components of
  gravitation, gravitomagnetism, and spacetime symmetries.}
\FullConference{
Advancing Astrophysics with the Square Kilometre Array\\
June 8-13, 2014\\
Giardini Naxos, Italy}

\begin{document}

\section{Introduction}
\label{sec:intro}

In November 1915, Albert Einstein published the final paper that
completed his theory of gravity and spacetime now known as general
relativity (GR), changing our view of gravity and spacetime for
ever. During its first fifty years, GR was considered a theoretical
tour-de-force but one with meager observational evidence. Only in the
1960s did technology usher in the remarkable field of experimental
gravity \citep{mtw73,will93}. Over the subsequent fifty years, GR has
passed all experimental tests in laboratories, in the Solar System and
in various stellar systems, particularly in binary pulsars
\citep{will14}.  Astrophysical observations using radio bands have
played one of the most important parts in the history of testing GR,
with the ability to make precise measurements and probe into the
vicinity of compact bodies, namely black holes (BHs) and neutron stars
(NSs).  Particularly, radio timing of binary pulsars has precisely
probed the gravitational properties of NSs and, for the first time,
tested the radiative properties of gravity, demonstrating that
gravitational waves exist and that compact binaries lose energy from
their emission at the rates predicted by GR
\citep{ht75,tfm79,ksm+06,wnt10,fwe+12,afw+13}. This confirmation and
precise characterisation of the radiative properties of gravity is not
only critical from the point of view of our understanding of
fundamental physics and many aspects of astrophysics, but it will also
open, via gravitational wave detectors, a whole new window on the
Universe.  The Square Kilometre Array (SKA), as the most powerful
member of the next generation of radio telescopes, will allow much
more precise radio timing of pulsars and will discover dozens of rare
relativistic binary systems. For this reason it will play a unique
role in many areas of experimental gravity.

Despite its success over the last century, GR still faces great
challenges on the research frontiers of modern physics.  First, it is
known to be incomplete, since it fails at the centre of BHs. Second,
as a non-renormalisable field theory, it appears to be incompatible
with quantum principles.  Reconciling gravity with quantum mechanics
stands out as one of the great challenges of fundamental physics
today. Furthermore, concepts like dark matter and dark energy pose
additional challenges for our understanding of gravity, and
modifications of gravity theories have been proposed to explain these
phenomena \citep{cl11}, like the Tensor-Vector-Scalar (TeVeS) theory,
which was advanced to explain dark matter \citep{bek04a}.

The important thing to us is that pulsar timing can be used to test
some of these theories.  Pulsar timing uses large radio telescopes to
record the times of arrival (TOAs) of pulsed signals, produced by the
rotation of the pulsar. These TOAs depend on the rotational and
astrometric parameters of the pulsar, dispersion in the interstellar
medium that the signals traverse, and the motion of the radio
telescope in the Solar System. If the pulsar is in a binary, the TOAs
also depend on the orbital dynamics of the binary, which are
determined by the underlying gravitational theory
\citep{wt81,dt92,ehm06}. Some theories of gravity, like a family of
TeVeS-like theories introduced in \citet{fwe+12}, result in orbital
dynamics that is sufficiently different from GR to be detected in the
timing of some binary pulsars. Therefore, precise timing measurements
of these binary pulsars can constrain, or altogether rule out, some of
these gravity theories.

For example, in the Jordan-Fierz-Brans-Dicke theory
\citep{jor59,fie56,bd61}, gravity is mediated by a scalar field,
$\varphi$, in addition to the canonical metric field,
$g_{\mu\nu}$. The variation of $\varphi$ in spacetime introduces
observable effects in gravitational experiments, like the
time-variation of the local gravitational constant $G$ and the
existence of gravitational dipole radiation in a binary. In a more
general class of scalar-tensor theories \citep{de93,de96b}, the
coupling strength of the scalar field $\varphi$ to matter depends on
the field itself $\varphi$ as $\alpha_0 + \varphi \beta_0$, where
$\alpha_0$ and $\beta_0$ fully characterise the theory.  Within some
of the $(\alpha_0, \beta_0)$ space, non-perturbative effects, called
``spontaneous scalarisation'', may happen in compact bodies like NSs
\citep{de93}, which can influence the orbital dynamics of a binary
system dramatically.  Similar effects are also possible within an
extended family of TeVeS-like theories \citep{fwe+12}.  With their
ability to probe strong-field effects, binary pulsars are ideal
laboratories with which to study these alternative gravitational
theories. With the non-detection of gravitational dipole radiation
from radio pulsar timing, binary pulsars have already severely
constrained the parameter spaces of these theories and provided the
most stringent tests of the strong-field\footnote{Compared to the
  coalescence of two compact objects (that will be the main source for
  ground-based gravitational wave detectors), two components of a
  binary pulsar are well separated, that might be prejudicially termed
  as ``weak field''. However, as shown by explicit calculations,
  ``strong-field effects'' associated with strongly self-gravitating
  NSs could have significant effects on the binary orbital dynamics
  \citep{de93}.} and radiative properties of gravity
\citep{fwe+12,afw+13,sta03,wex14}.

\begin{figure}
\centering
\includegraphics[width=15cm]{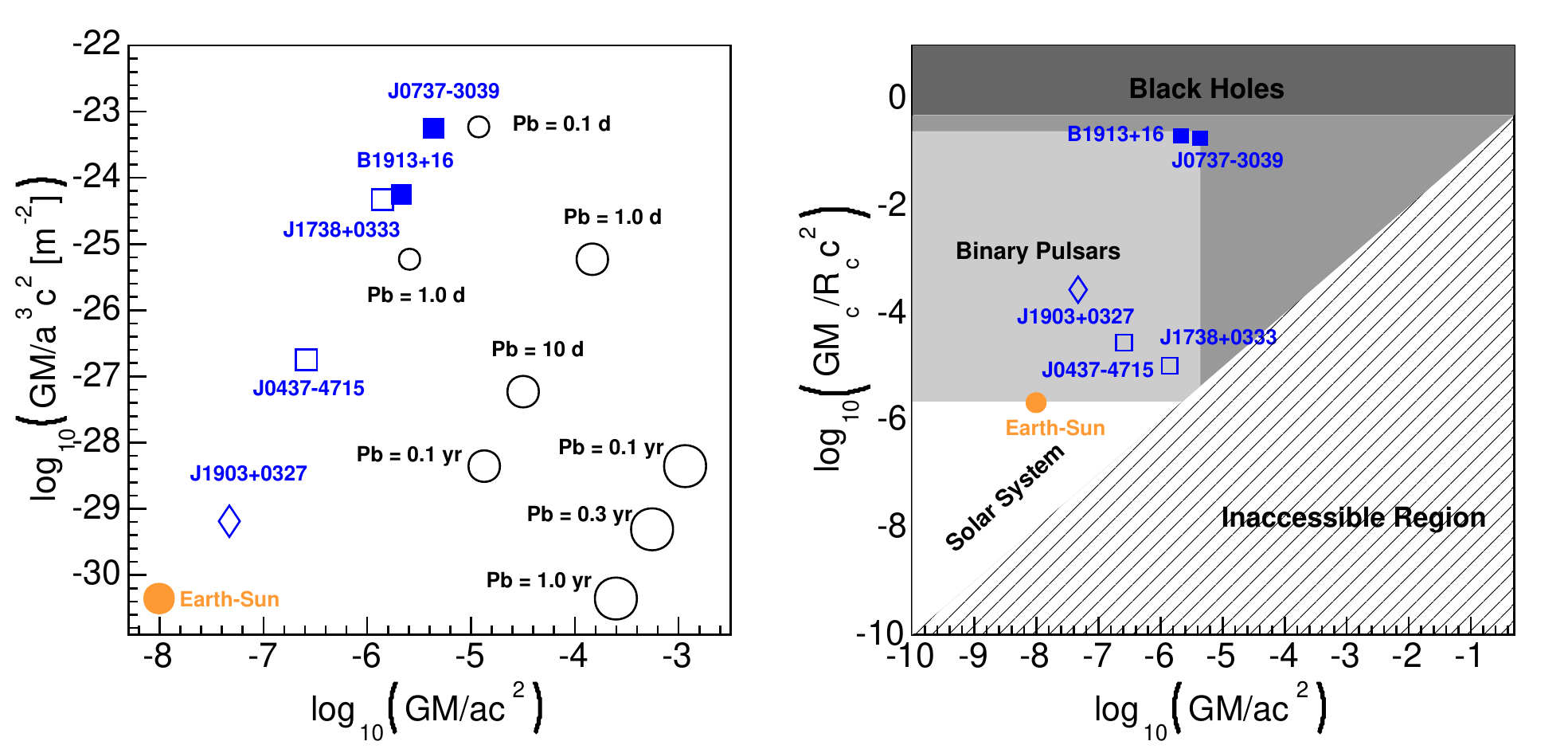}
\caption{{\it Left:} Parameter space for quantifying the strength of a
  gravitational field \citep{psa08}. The horizontal axis measures the
  gravitational potential and the vertical axis measures the spacetime
  curvature of an orbit with a semimajor axis, $a$, and a total mass,
  $M$ (here $G$ is the gravitational constant, and $c$ is the light
  speed).  Solid and empty squares are known neutron star -- neutron
  star (NS-NS) and neutron star -- white dwarf (NS-WD) binaries,
  respectively. The diamond is a known neutron star -- main sequence
  star binary.  Circles from large to small are pulsar -- Sgr A* BH,
  pulsar -- 5000\,$M_\odot$ BH, and pulsar -- 10\,$M_\odot$ BH
  binaries with their assumed orbital periods aside. {\it Right:} The
  compactness of the gravitating companion star versus the compactness
  of the orbit; $M_c$ and $R_c$ are the mass and the radius of the
  companion star, respectively \citep{kbc+04}.
  \label{fig:grav}}
\end{figure}

\subsection{The role of the Square Kilometre Array}

The Square Kilometre Array will have a major impact on radio
astronomy.  The SKA Phase~1 (SKA1) will feature roughly half of its
total collecting area within a dense core, in the cases of SKA1-LOW
and SKA1-MID. By synthesising a coherent sum of these elements, it
will be possible to deeply search for new relativistic binary pulsars,
over the whole sky, and to subsequently time them with great precision
using the full SKA1-MID array ({\it i.e.} including the longer
baselines).  The SKA Phase 2 (SKA2) will provide another major leap in
sensitivity, and thus also in timing precision.  Both phases of the
SKA will significantly advance experimental tests of gravitational
theories to a new level \citep{kbc+04}.

This will happen through two avenues.  On the one hand, SKA1-MID and
SKA2 will strongly improve the quality of the observations of the most
interesting binaries currently known ({\it e.g.}, PSR~J0737$-$3039
\citep{bdp+03,lbk+04,ksm+06}) and triple systems ({\it e.g.},
PSR~J0337+1715 \citep{rsa+14}), in turn dramatically improving the
quality of the existing tests and enabling proposed new tests.
Compared with the current status quo, SKA1-MID will improve the timing
precision for southern pulsars by about an order of magnitude, and
SKA2 will improve it by up to two orders of magnitude.  Different
pulsars will gain different improvements in precision, depending on
their sky locations, pulse characteristics, timing stabilities, and so
on.

For precision pulsar timing, raw sensitivity is one of the most
important factors.  High cadence may also be needed for some
experiments, in which case sub-arraying will be needed.  Typically,
TOAs for a particular pulsar are collected every few weeks, though in
come cases a much higher cadence is necessary, {\it e.g.} to focus on
orbital conjunction in order to measure the Shapiro delay.
Sub-arraying can also help if many sources need to be timed, and if
reduced sensitivity is acceptable.  Though SKA1-MID will provide the
highest-precision TOAs, regular measurements with SKA1-LOW may also be
needed to track the time-variable interstellar propagation effects
that also influence the TOAs.  In all cases, coherently dedispersing
the data is essential for achieving the maximum timing precision and
accuracy.  Such timing observations can already begin in an early
phase of SKA1 (say, with 50\% of the collecting area available),
though they will naturally obtain less precise TOAs than with the
complete array.  For some sources 50\% sensitivity will be sufficient,
while for others certain relativistic effects may only become clearly
detectable once the full array is in place.

The SKA will also {\it discover} new relativistic systems with which
to carry out similar and completely novel tests.  With a larger field
of view and high sensitivity, the SKA is more efficient in finding new
pulsars.  The larger sensitivity of the SKA means that a shorter
integration time is needed to achieve the required signal to noise
ratio. Consequently, in an ``acceleration search'' with an assumed
constant orbital acceleration \citep{lk04}, the smearing due to
varying acceleration in an actual binary will be reduced, making the
SKA much more effective than current telescopes at finding
relativistic binaries.  Simulations show that the SKA1 can discover a
total of about 10,000 normal pulsars and perhaps as many as 1,800
millisecond pulsars (MSPs), with SKA1-LOW surveying the sky with the
Galactic latitude $|b| \geq 5^\circ$, and SKA1-MID surveying the sky
with the Galactic latitude $|b| \leq 10^\circ$ \citep{kbk+14}. The
SKA2 will discover even more pulsars, and it is possible that among
them there will be pulsar -- black hole (PSR-BH) binaries
\citep{kbc+04,kbk+14}. The discovery of even a single tight PSR-BH
system will open a new era of studying BH physics with great
precision, including possible tests of the ``cosmic censorship
conjecture'' and the ``no-hair theorem''
\citep{de98,wk99,kbc+04,liu12,lwk+12}.  Figure~\ref{fig:grav} shows
the vast unprobed regions in the gravity sector that can be probed
with different kinds of PSR-BH systems.

\section{Relativistic Binaries}
\label{sec:binary}

Pulsar timing takes advantage of the tremendous rotational stability
of pulsars, allowing us to treat them as free-falling clocks.  As a
pulsar is monitored over months and years, an ephemeris is calculated
which accounts for every rotation of the pulsar and can be used to
predict future pulse TOAs.  A pulsar ephemeris always incorporates the
spin frequency and generally at least one spin derivative accounting
for rotational energy loss, plus astrometric parameters.  Binary
pulsar orbits require a further translation of reference frame,
accomplished for most binary systems by the five Keplerian parameters
describing an eccentric Keplerian orbit. In order to account for
relativistic effects beyond a Keplerian orbit, a set of
theory-independent ``post-Keplerian'' (PK) parameters are introduced
\citep{dd86,dt92} (see Table~\ref{tab:PK} for an incomplete collection
of the most important PK parameters).  In practice, PK parameters
could be contaminated by astrophysical effects other than relativistic
effects due to gravity. For example, the observed orbital decay
parameter, $\dot P_b$, has ``kinematic'' contributions, which result
from the proper motion of the system and the difference of Galactic
accelerations between that system and the Solar System Barycentre.
Here unless otherwise stated, we assume that non-gravitational
contributions are properly corrected or demonstrably negligible in
effect. If so, in the case of two well-separated masses with
negligible spin contributions, the PK parameters are functions of the
well-measured Keplerian parameters, the component masses, the equation
of state (EOS) of stellar matters, and the parameters describing the
gravitational theory \citep{dt92} (in GR, the internal constitution of
the star is irrelevant to a high ``post-Newtonian'' (PN) order, only
the total masses matter --- this is the ``effacement'' property to be
discussed in section 3).

\begin{figure}[t]
  \centering
  \includegraphics[width=13cm]{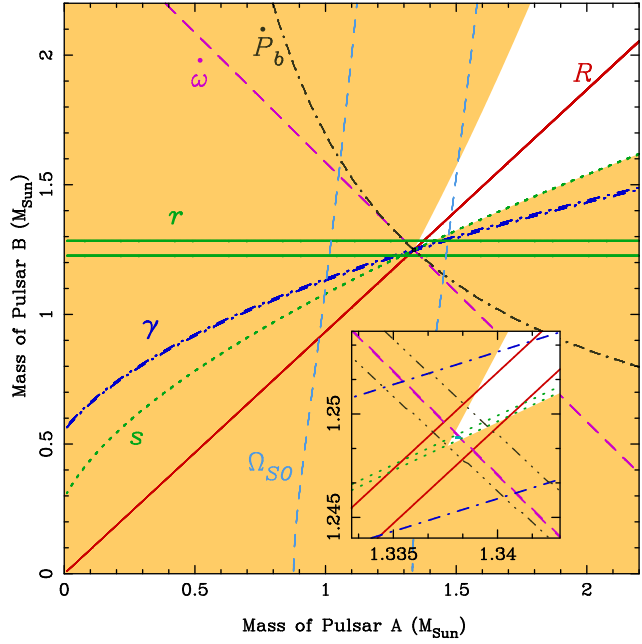}
  \caption{The mass-mass diagram of PSR~J0737$-$3039A/B, also known as
    the Double Pulsar (Kramer et al., in prep.). In the figure the
    underlying gravitational theory is assumed to be GR.  Shaded
    regions are forbidden by the individual mass functions because the
    sine of the orbital inclination must be $\leq 1$.  The inset is an
    expanded view of the region of principal interest, where the
    intersection of seven curves at one point within measurement
    uncertainties proves the existence and uniqueness of a
    solution. For more details, see \citet{ksm+06} and
    \citet{bkk+08}.\label{fig:J0737}}
\end{figure}

Measuring two PK parameters, we can use a specific theory of gravity
to determine the component masses of the system; these are important
to study stellar evolution theories and, in some cases, to constrain
EOS \citep{wxe+14}.  If more PK parameters are measured, then the
theory can be tested by a self-consistent argument --- using the
masses derived in the first stage one should be able to predict the
subsequent PK measurements.  Another way of viewing this is the
so-called ``mass-mass diagram'' (see Figure~\ref{fig:J0737} for a
GR-based example for PSR~J0737$-$3039; Kramer et al., in prep.). For a
gravitational theory to pass the test(s), all curves in the diagram
should intersect in some region, {\it i.e.}, the theory must be able
to describe the component masses in a self-consistent way
\citep{tw82}.

\begin{table}[t]
  \caption{The most important PK parameters that could be used in
    pulsar timing of binaries \citep{dd86,dt92,wk99,lk04,ehm06}.  In
    practice, for a specific binary pulsar, only some PK parameters
    are measured, depending on the characteristics of the pulsar
    timing experiment.\label{tab:PK}} \centering
  \begin{tabular}{ll}
    \hline\hline
    Parameter &  \\
    \hline
    $\dot\omega$ & time derivative of the longitude of periastron
    $\omega$ \\
    $\gamma$ & amplitude of the Einstein delay\\
    $\dot P_b$ & time derivative of the orbital period $P_b$ \\
    $r$ & range of the Shapiro delay\\
    $s$ & shape of the Shapiro delay \\
    $\Omega_{\rm SO}$ & precession rate of the pulsar spin \\
    $\delta_\theta$ & mismatch in eccentricities (see text) \\ 
    $\dot e$ & time derivative of the orbital eccentricity $e$ \\
    $\dot x$ & time derivative of the projected semimajor axis of the
    pulsar orbit $x$\\
    $\ddot \omega$ & second time derivative of the longitude of
    periastron  \\
    $\ddot x$ & second time derivative of the projected semimajor axis of the
    pulsar orbit \\
    \hline
  \end{tabular}
\end{table}

The current state-of-the-art binary in this area is the Double Pulsar
(PSR J0737$-$3039A/B) \citep{bdp+03,lbk+04}, which enables five tests
of GR. Figure~\ref{fig:J0737} shows its mass-mass diagram, where seven
curves corresponding to seven mass constraints intersect at one point
\citep{ksm+06,bkk+08,kw09}, indicating that GR provides a
self-consistent description of all of its timing measurements.  The
Double Pulsar resides in the southern sky and will be an important
target for the SKA. The timing precision of the SKA will enable tests
of relativistic effects beyond their first-order PN approximations
\citep{kw09}. Measuring higher-order PN effects is important since it
allows the investigation of the effects from pulsar spin
\citep{bo75,ds88,kw09}. The fractional effects due to the next PN
order\footnote{Effects due to the PN orders are those that are
  suppressed by the ratio of the orbital velocity, $v$, to the light
  speed, $c$, compared with those due to Newtonian gravity
  \citep{will14}. It is denoted as ``$n$PN'' if the suppression factor
  is ${\cal O}(v^{2n}/c^{2n})$.}  on the advance rate of periastron,
namely $\dot\omega^{\rm 2PN} / \dot\omega^{\rm 1PN}$, and on the
orbital decay, namely $\dot P_b^{\rm 3.5PN} / \dot P_b^{\rm 2.5PN}$,
are both at the order of ${\cal O}(10^{-5})$. Simulations showed that
by obtaining a timing variance below $\sim 5\,\mu$s for the Double
Pulsar we will reach the precision required to detect higher-order
effects \citep{kw09}; this is very likely with SKA1-MID, scaling from
current observations.  The SKA2 will have sub-$\mu$s timing precision
for the Double Pulsar (assuming its pulse profile shape can still be
well modelled), raising the prospect of highly precise measurements of
these higher-order parameters.

Besides the orbital dynamics of two point masses, spin contributions
may play an important role in the observations of pulsar binaries
\citep{bo75}. Due to the curvature of the spacetime produced by the
companion star, the rotation axis of a freely falling object (here the
pulsar) suffers a precession with respect to a distant observer; this
effect is known as ``geodetic precession''. Because of it, over time
different emitting regions of the magnetosphere will be visible from
the Earth, thus causing a secular change in the observed pulse
profile. This effect has been observed in some tight binary pulsars,
for example, in PSRs~B1913+16 \citep{wrt89,kra98}, B1534+12
\citep{sta04,fst14}, J1141$-$6545 \citep{mks+10}, J1906+0746
\citep{lsf+06} and J0737$-$3039B \citep{pmk+10}. For PSRs~B1534+12 and
J0737$-$3039B we even have direct measurements of the precession
rates, which match the GR predictions \citep{sta04,fst14,bkk+08}.  The
SKA1-MID will have comparable sensitivity to the Arecibo telescope,
meaning that significant improvement in precession measurements of the
pulsars that are in the Arecibo sky will await the SKA2.  However,
precessing pulsars not visible to the Arecibo telescope will be prime
targets for SKA1-MID.  For example, in the Double Pulsar, the
precession rate of the B pulsar can be measured by modelling the
orientation of its magnetosphere as it causes ``flickering'' eclipses
of pulsar A \citep{bkk+08}.  The rapidly changing eclipse-period flux
density will be much better measured by both phases of the SKA,
leading to a more stringent test of the precession rate's agreement
with GR.

Since in GR the total angular momentum must be conserved up to 2PN
order, a change in the direction of the pulsar spin must be
compensated by a change in the direction of the orbital angular
momentum, contributing a variation to the projected semimajor axis of
the orbit ($\dot x \neq 0$).  This spin-orbit coupling is also known
as the Lense-Thirring effect or frame-dragging effect, and it is a
small effect that is numerically at 2PN order for slowly rotating
neutron stars, whose measurement requires a high timing precision
\citep{ds88}.  However, measuring it is important since it offers the
potential to measure the moment of inertia of the pulsar
\citep{ds88,oco04}, of great interest for the EOS of cold dense
nuclear matter \citep{wxe+14}.
  
For the Double Pulsar, the recycled pulsar (pulsar A) dominates the
contribution to the Lense-Thirring precession, because it is a faster
rotator.  The effect on $\dot{x}$ will be almost impossible to
measure, not only because in this case there is a close alignment of
the orbital angular momentum and the spin of pulsar A \citep{fsk+13},
but also because the system is viewed almost edge-on
\citep{ksm+06}. However, the SKA Galactic census will discover many
NS-NS systems (scaling from the current population, we expect 100 from
the SKA1 and 180 from the SKA2 \citep{kbk+14}), and there may well be
binaries that will enable the direct measurement of $\dot{x}$. In
order to achieve the goal, a tight binary with good timing precision
is needed. In addition, the misalignment angle between the pulsar spin
and the orbital norm should be reasonably large, {\it i.e.}, geodetic
precession should be observable.  This requires a relatively large
``supernova kick velocity'' for the second-formed compact object ---
in other words, an iron-core-collapse supernova, instead of an
electron-capture supernova as likely happened in the Double Pulsar
\citep{fsk+13}. In such misaligned binaries, the frame dragging
introduces nonlinear evolution in time in the longitude of periastron
and its projected semimajor axis; thus one might be able to extract
$\ddot\omega$ and $\ddot x$ from radio timing. This would impose
useful constraints on the system's geometry \citep{wk99}.

The Lense-Thirring effect also contributes to $\dot\omega$ with a
magnitude comparable to that of $\dot\omega^{\rm 2PN}$
\citep{ds88,kw09}.  For the Double Pulsar, this can in principle be
extracted from the advance rate of periastron. To do this we must
determine the component masses very precisely using two other precise
PK parameters. One of them (the shape of Shapiro delay, $s$) already
possesses the required precision (Kramer et al., in prep.). For the
other one, $\dot{P}_b$, the orbital decay due to the emission of GW
waves, we will need to estimate the distance to the pulsar carefully,
in order to estimate precisely the kinematic corrections to it. This
can only be done with the timing precision provided by the SKA. Again,
if this Lense-Thirring contribution to $\dot\omega$ can be estimated
precisely, we will then be able to estimate the moment of inertia of
pulsar A with similar precision. Because this measurement will be done
for a pulsar with an exquisitely well-determined mass, it will be a
very important constraint on the EOS for dense matter
\citep{ls05,kw09,wxe+14}.

The SKA will also confirm general-relativistic orbital deformation for
the first time.  \citet{dd86} used three different eccentricities to
describe the timing of a pulsar. Two parameters were introduced in the
pulsar timing formula, namely $\delta_\theta$ and $\delta_r$, where
the former is in principle an observable \citep{dt92}. However, it
only adds a tiny periodic variance to the TOAs and the signal has a
strong correlation with the Einstein delay that is caused by the
transverse Doppler effect and the gravitational redshift, its
amplitude is $e^2 / \sqrt{1-e^2}$ in binaries of mass ratio near
unity, where $e$ is the eccentricity of the orbit.  Therefore, we will
need highly eccentric orbits to ``amplify'' the effect, and highly
relativistic orbits with large periastron advance rate to break the
degeneracy with the Einstein delay. The SKA is likely to find such a
system in the southern sky and time it with high precision. We
simulate TOAs with orbital parameters similar to that of PSR~B1913+16
\citep{ht75,wnt10} with the {\tt TEMPO}
software\footnote{http://tempo.sourceforge.net}. It is assumed that 60
TOAs with a precision $\sim 1\,\mu$s are obtained per month with the
SKA1.  After 10 years of observation, the $\delta_\theta$ parameter
can be determined to a precision $\sim 10\%$. A more eccentric (or a
more relativistic) binary can achieve a better test. These results
support earlier simulations done with the Double Pulsar \citep{kw09}.

Where parameters in the timing model can be measured to high precision
by other means, they can be restricted to an allowed range in the
timing fit. This can lead to significantly reduced uncertainties for
the remaining parameters in the presence of covariances. The most
common source of external information is the estimation of position,
proper motion and annual parallax using Very Long Baseline
Interferometry (VLBI). The use of VLBI can improve gravitational tests
when applied to pulsar binaries.  For instance, astrometric
information provided by VLBI was used to calculate the kinematic
contamination of the measured orbital period derivative of the Double
Pulsar, improving the potential accuracy of tests of gravitational
radiation to the 0.01\% level \citep{dbt09}.  In a second example, the
previously unmeasured Shapiro delay for PSR J2222$-$0137 was precisely
ascertained after the position, proper motion, and parallax were fixed
using VLBI information \citep{dbl+13}. Since VLBI measures motions on
the plane of the sky, it has different ``blind spots'' with respect to
pulsar timing; VLBI has no ``blind spots'' to parallax at the ecliptic
pole and proper motion in the ecliptic plane.  With current VLBI
instruments, precisions of 10 micro-arcseconds for parallax and 10
micro-arcseconds per year for proper motion can be reached
\citep{dbl+13}.  The SKA1 will allow a similar leap in VLBI precision
as it will provide in the timing precision, and so VLBI using the SKA1
will be able to deliver better results for parallax and proper motion
than timing in most cases for nearby pulsars \citep{pgr+14}.  These
measurements will be very important to subtract the Shklovskii
contribution (which usually dominates the error budget), and the
differential Galactic rotation, and vertical acceleration in the
Galactic potential (which also depend on relatively well-constrained
Galactic models), thus measure the intrinsic orbital period change to
high precision.

\section{Gravitational Dipole Radiation and Equivalence-Principle Tests} 
\label{sec:gw:sep}

One of the characteristics of GR is the property of ``effacement''
\citep{dam87}. It states that the finite extents of two bodies in a
binary are largely irrelevant to their orbital dynamics as long as
they are well separated.  Effacement is deeply rooted in the strong
equivalence principle (SEP) \citep{will93}.  However, in alternative
theories of gravity, the SEP is generally violated, and the strong
gravitational fields associated with a NS may have measurable impact
on the orbital dynamics of the binary and the trajectory of light
propagation \citep{will93,de92a,de93}.  Figure~\ref{fig:grav} shows
various binary systems, with the gravitational field strength
characterised by the compactness and curvature of the orbit and the
compactness of the companion star.

\subsection{Gravitational dipole radiation and time-variation of the
  gravitational constant}

\begin{figure}
  \centering
  \includegraphics[width=10cm]{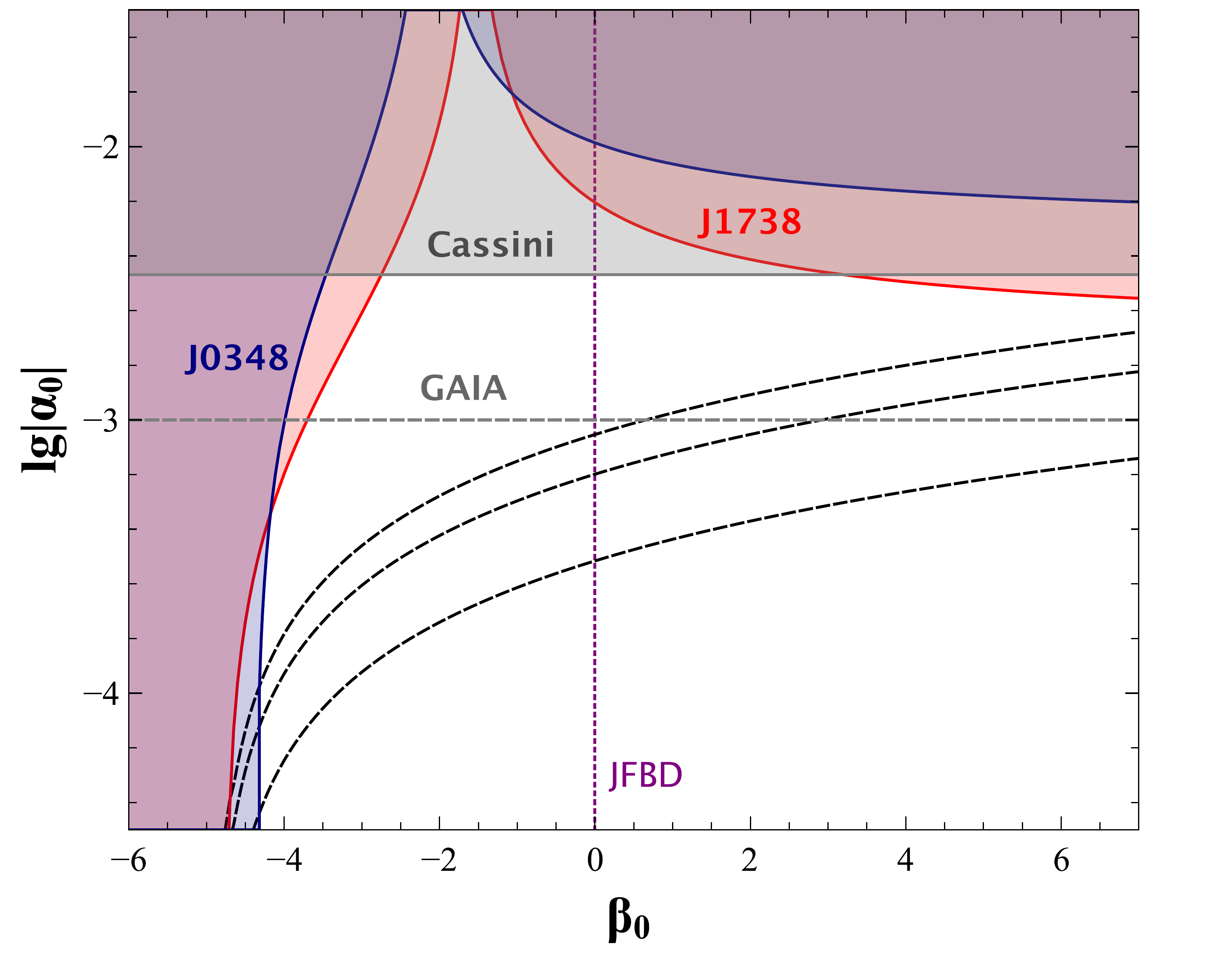}
  \caption{Constraints on the parameter space of scalar-tensor
    theories (figure courtesy of N. Wex). Coloured regions are
    excluded by current Solar System and pulsar experiments: Cassini
    spacecraft \citep{bit03}, PSR~J1738+0333 \citep{fwe+12}, and
    PSR~J0348+0432 \citep{afw+13}. The EOS is chosen as MPA1. Black
    dashed lines are based on simulations for a MSP-BH system with
    $(M_{\rm BH}, M_{\rm MSP}) = (10\,M_\odot, 1.4\,M_\odot)$, $P_b =
    5$\,d, and $e=0.8$; the constraints from top to bottom are based
    on: 10 years with a 100-m class radio telescope, 5 years with the
    FAST telescope, and 5 years with the SKA \citep{lewk14}. The
    vertical dashed purple line ``JFBD'' indicates the
    Jordan-Fierz-Brans-Dicke gravity. The horizontal dashed grey line
    ``GAIA'' indicates the limit expected from near-future
    Solar-System experiments, foremost from the astrometric satellite
    GAIA.
    \label{fig:scalartensor}}
\end{figure}

A possible violation of the SEP could be identified by the occurrence
of gravitational dipole radiation. The dipole radiation is at lower PN
order than the canonical quadrupole radiation predicted by GR, thus in
principle it can dominate over the quadrupole mode
\citep{de92a,de93,will93,will14}.  It carries away the gravitational
energy of the orbit more rapidly than that in GR, collapsing the
binary more quickly, and modifying the waveform of gravitational
radiation that can be detected by future ground-based GW
detectors. The leading change in the orbital period of a binary is
given by the dipolar contribution with a form $\dot P_b^{\rm dipole}
\propto (\alpha_1 - \alpha_2)^2$, where $\alpha_i$ ($i=1,2$) is the
effective coupling strength between body $i$ and the fields associated
with the dipole moment. In the best-studied gravitational theory of
this type, the scalar-tensor theory \citep{de92a,will93}, we generally
have $\alpha_{\rm NS} \neq \alpha_{\rm WD}$ in a NS-WD binary (see the
right panel of Figure~\ref{fig:grav}). Therefore, NS-WD binaries are
the ideal laboratories for testing these theories and, if they are
correct, searching for the existence of gravitational dipole
radiation, which if detected would falsify GR.

In these NS-WD binaries, the combined radio observation of the pulsar
and optical observation of the WD can give rise to precise
measurements of two component masses \citep{akk+12,afw+13,wex14}.  The
mass measurements can be used to predict the amount of gravitational
radiation within a specific theory, and this in turn is confronted
with the radio timing observation of the orbital decay caused by
gravitational damping. As mentioned, the dipolar radiation can in
principle be dominant over the quadrupolar radiation, therefore,
agreement between the observed $\dot P_b$ and the quadrupole formula
in GR stringently constrains the dipolar radiation contribution, as in
the cases of PSRs~J1738+0333 \citep{fwe+12} and J0348+0432
\citep{afw+13}.

In Figure~\ref{fig:scalartensor}, the constraints from those two
binaries are plotted in the $\alpha_0$-$\beta_0$ parameter space
mentioned in section 1. The best constraint from the Cassini
spacecraft in the Solar System is also plotted. The limits from two
binary pulsars stringently constrain the parameter space of
scalar-tensor theories, especially when the $\beta_0$ parameter is
negative. The reason for this is that, as showed by \citet{de93}, in
this region NSs develop non-perturbative strong-field effects like
``spontaneous scalarisation'' that are absent in the weak-field
experiments in the Solar System; such effects would cause a very large
increase in the emission of dipolar gravitational waves. The more
massive the pulsar the less negative $\beta_0$ has to be for the
spontaneous scalarisation to appear.  For this reason, the timing of
PSR~J0348+0432, a two-solar-mass NS in a relativistic orbit where the
orbital decay has been measured with some precision, introduces
uniquely stringent constraints on the allowed range of $\beta_0$
\citep{afw+13}.

In Jordan-Fierz-Brans-Dicke theory and (from a broader viewpoint) in
the generic scalar-tensor theories, the scalar field plays the role of
local gravitational constant, implying that the latter may vary with
space and time \citep{de92a,will93}. The possible time-variation in
the gravitational constant contributes to the decay of a binary orbit
\citep{dgt88,wex14}. The extra contribution can be constrained by the
measurement of $\dot P_b$ for binary pulsars by assuming a null
contribution from the dipole radiation.  In general, however, a
varying gravitational constant is accompanied by the dipole radiation
\citep{lwj+09}.  With more than one binary pulsar, one can also
conduct a joint constraint simultaneously on both the dipole radiation
and the time-variation in the gravitational constant, utilising the
fact that these two effects have different dependence on the orbital
period of the binary \citep{lwj+09}. The latest constraint from
pulsars on the time-variation of the gravitational constant
\citep{fwe+12}, which relies on VLBI \citep{dvtb08} and timing
\citep{vbs+08} of PSR~J0437$-$4715 and J1738+0333, is already
comparable to the best constraint from the Solar System experiments
\citep{will14}. Moreover, pulsar tests are sensitive to strong-field
effects on $\dot{G}$ \citep{wex14}.

These examples already show how important proper subtraction of the
contributions from Galactic acceleration and the Shklovskii effect is,
and how crucial VLBI is for reducing uncertainties in distances and
proper motions. Furthermore, in the near future a more accurate
understanding of the Galactic potential will also be available ({\it
  e.g.}, by the GAIA satellite\footnote{http://sci.esa.int/gaia/}),
further improving the estimates of the kinematic corrections of $\dot
P_b$.

The next generation of 30-m class optical telescopes, for examples,
the Thirty Meter Telescope (TMT)\footnote{http://www.tmt.org/}, the
Giant Magellan Telescope (GMT)\footnote{http://www.gmto.org/}, and the
European Extremely Large Telescope
(E-ELT)\footnote{http://www.eelt.org.uk/}, may also significantly
advance the optical observations of WDs, improving their mass and
distance estimates; which will soon limit the precision of some of the
current gravity tests \citep{fwe+12}.

In short, in the SKA era, pulsar observations will improve
significantly and maintain their unique importance in testing
alternative theories of gravity, but will benefit extensively from
inputting other techniques like VLBI and optical observation.  They
will be crucial for tests of some extended gravitational theories that
incorporate dark matter and dark energy as gravitational effects.

If any relativistic PSR-BH binaries are discovered, they will also be
promising test beds for gravitational dipolar radiation because the
no-hair theorem also applies to scalar-tensor theories of gravity;
this implies that BHs have no scalar charge\footnote{However, see
  \citet{hr14} for ``Kerr BHs'' with scalar hair after a complex,
  massive scalar field, minimally coupled to GR, is included.}.  A
consequence of this is that any radiative test that constrains
$(\alpha_{\rm NS} - \alpha_{\rm BH})$ will therefore constrain
$\alpha_{\rm NS}$ directly \citep{de98}. The constraints from
simulations of a hypothetical PSR-BH system ($M_{\rm BH} = 10\,{\rm
  M}_\odot$, $P_b = 5$\,day and $e=0.8$) are depicted as dashed lines
in Figure~\ref{fig:scalartensor} \citep{lewk14}. Five years of weekly
observations with the SKA will constrain $|\alpha_0|$ down to the
order of ${\cal O}(10^{-4})$ \citep{liu12,wle+13}, improving upon the
expected constraint from the GAIA satellite.  Worthy to note that,
unlike NS-WD binaries, these tests with PSR-BH binaries will also
constrain the parameter space with $\beta_0 \sim -1$.

\subsection{Equivalence-principle violation and its effects on orbital
dynamics}

A possible violation of the SEP has also been investigated by tracing
the orbital dynamics of small-eccentricity NS-WD binaries
\citep{ds91}. The differential free fall rates of the NS and the WD in
the potential of the Milky Way will induce a characteristic time
evolution of the orbital eccentricity vector \citep{ds91}.  The SEP
violation tends to polarise the Laplace-Runge-Lenz vector of the orbit
towards the direction of the Galactic acceleration. By using
statistical studies on the eccentricity distribution of NS-WD binaries
with a probabilistic assumption \citep{wex00}, the dimensionless SEP
violation parameter (the Nordtvedt parameter), $\eta$, was constrained
to a comparable magnitude to that from the tests in the Solar System
\citep{sfl+05,gsf+11,will14}.  A direct test of SEP with binary
pulsars that uses the time-variations in orbital parameters was also
developed \citep{fkw12}. It does not rely on statistical or
probabilistic assumptions.

The recently discovered triple system, PSR~J0337+1715 \citep{rsa+14},
provides an opportunity to improve the SEP test by many orders of
magnitude \citep{fkw12}.  In this system, an outer WD orbits an inner
NS-WD binary in less than one year.  Usually, in the tests of SEP
violation, the Galactic acceleration exerted on the binary is of the
order ${\cal O}(10^{-10}\,{\rm m\,s}^{-2})$, which is the force that
would ``polarise'' the inner orbit in the case of SEP violation. In
PSR~J0337+1715, the outer WD provides a ``polarising'' acceleration
that is more than one million times larger than the Galactic
acceleration on the inner pair.  As a rough estimate, we use the
acceleration in Eq.~(4) of \citet{ds91}, by replacing the relativistic
rate of advance of periastron for the inner binary with the {\it
  orbital} angular velocity of the outer orbit (also see Eq.~(8.18) in
\citet{will93}).  Based on the current measurement precision of the
orbital eccentricities in the triple system, a possible upper limit of
the SEP violation parameter is estimated, $|\Delta| \lesssim 4 \times
10^{-7}$, which is $10^4$ times better than the current pulsar
constraint (N. Wex, private communication). Due to this pulsar's
detectability with the Arecibo telescope, observations with SKA1-MID
will mostly improve the SEP-violation test via the accumulation of
more data\footnote{In principle, SKA1-MID is better due to less radio
  frequency interference (RFI), and much wider bands than that of the
  Arecibo telescope.}, but the SKA2 will improve the timing precision
by a factor of ten, which should improve the test by a corresponding
factor.

It was estimated in \citet{rsa+14} that there could be $\lesssim 100$
such systems with MSPs residing in the Galaxy. Since the SKA will
discover almost all visible pulsars in the Galaxy \citep{kbk+14}, it
is likely that it will discover other triple systems like
PSR~J0337+1715, and it is possible that the SKA might even discover a
triple system consisting of an inner NS-WD close binary and where the
outer star is another NS.  Such a system could be (again) many orders
of magnitude more sensitive to SEP violation effects than
PSR~J0337+1715.  The reason is that the SEP violation effects are
proportional to $(\alpha_1^{\rm inner} - \alpha_2^{\rm inner}) \cdot
\alpha_3^{\rm outer}$ \citep{fkw12}, where $\alpha_{1,2}^{\rm inner}$
are the couplings of the inner bodies to the scalar field and
$\alpha_3^{\rm outer}$ is the coupling of the outer body to the scalar
field. If the outer body is a WD, then $\alpha_3^{\rm outer} \simeq
\alpha_0$ ($\alpha_0$ is coupling in the weak field), but if it is a
NS, then $\alpha_3^{\rm outer}$ could, according to some alternative
theories of gravity, be many orders of magnitude larger than
$\alpha_0$.  For such theories, the predicted SEP violation effects
would be roughly ``$\alpha_{\rm NS}/\alpha_0$'' times larger than
their prediction for PSR~J0337+1715, therefore making the system much
more sensitive to the SEP violation predicted in these theories.  The
non-detection of SEP violation in such a system would provide a
stringent constraint on alternative theories of gravity that predict
such SEP violation.

\subsection{Local Lorentz invariance and local position invariance of
  gravity} 

Besides the universality of the free fall, the other two aspects of
the Einstein equivalence principle \citep{will14}, namely local
Lorentz invariance (LLI) and local position invariance (LPI) of
gravity, can also be tested with pulsar timing experiments. Scenarios
with LLI violation have recently raised great interests in the
gravitational community, for examples, in the Einstein-\ae{}ther
theory and the Ho{\v r}ava-Lifshitz theory. Two generic frameworks for
testing deviations from GR are the parametrised post-Newtonian (PPN)
formalism \citep{will93,will14} and the standard-model extension (SME)
\citep{bk06}, both of which contain parameters for LLI violation.

Generically, violation of LLI and LPI leads to modifications of the
orbital dynamics of binary pulsars and the spin evolution of solitary
pulsars in characteristic ways
\citep{de92b,bk06,wk07,sw12,sck+13,shao14,sw13}. The LLI-violating
modifications of the orbital dynamics of a binary pulsar introduce
time-variations in the orbital eccentricity and orbital inclination,
plus an extra contribution to the periastron advance rate, thus
resulting in changes in the PK parameters, $\dot e$, $\dot x$, and
$\dot \omega$ \citep{de92b,bk06,sw12,shao14}.  The violation of LLI
also leads to spin precession of a solitary pulsar with respect to a
fixed direction\footnote{If there exists a preferred frame, the fixed
  direction will be the direction of the ``absolute'' velocity of the
  binary with respect to this preferred frame.} \citep{sck+13,shao14},
while the violation of LPI leads to spin precession of a solitary
pulsar around its ``absolute'' acceleration towards the Galactic
centre \citep{sw13}.  They both change our view on the emission region
of a pulsar, and result in characteristic changes versus time in the
pulse profiles.  Interestingly, modifications of the orbital dynamics
and spin evolution are related and controlled by one set of parameters
in both PPN and SME frameworks. These modifications can be jointly
constrained and cross-checked. The current best constraints on the LLI
of gravity and LPI of gravity are from pulsar experiments
\citep{sfl+05,sw12,sck+13,sw13,shao14}.  It was shown that these
constraints are proportional to the timing precision of binary
pulsars. With the SKA, tests will be improved dramatically with better
sensitivities. Simulations show that, if the signal to noise ratio of
pulse profiles improves by a factor of 10 with the SKA, then the
limits on Lorentz-violating parameters improve the same factor over
the current best ones with solely 10-year observations (say, from 2020
to 2030). The acquisition of one stable profile biweekly is assumed
for two MSPs in \citet{sck+13}. If these SKA observations are properly
combined with the 20-year pre-SKA observations (say, from 2000 to
2020), the improvement will be a factor of $\sim50$. In addition, new
stable MSPs from the Galactic census will be very helpful in breaking
the degeneracies of parameters in the test, due to the vectorial
and/or tensorial field condensations in the spacetime with such
violations \citep{shao14}.

\section{Pulsar -- Black Hole Binaries}
\label{sec:bh}

\begin{figure}
  \centering
  \includegraphics[width=12cm]{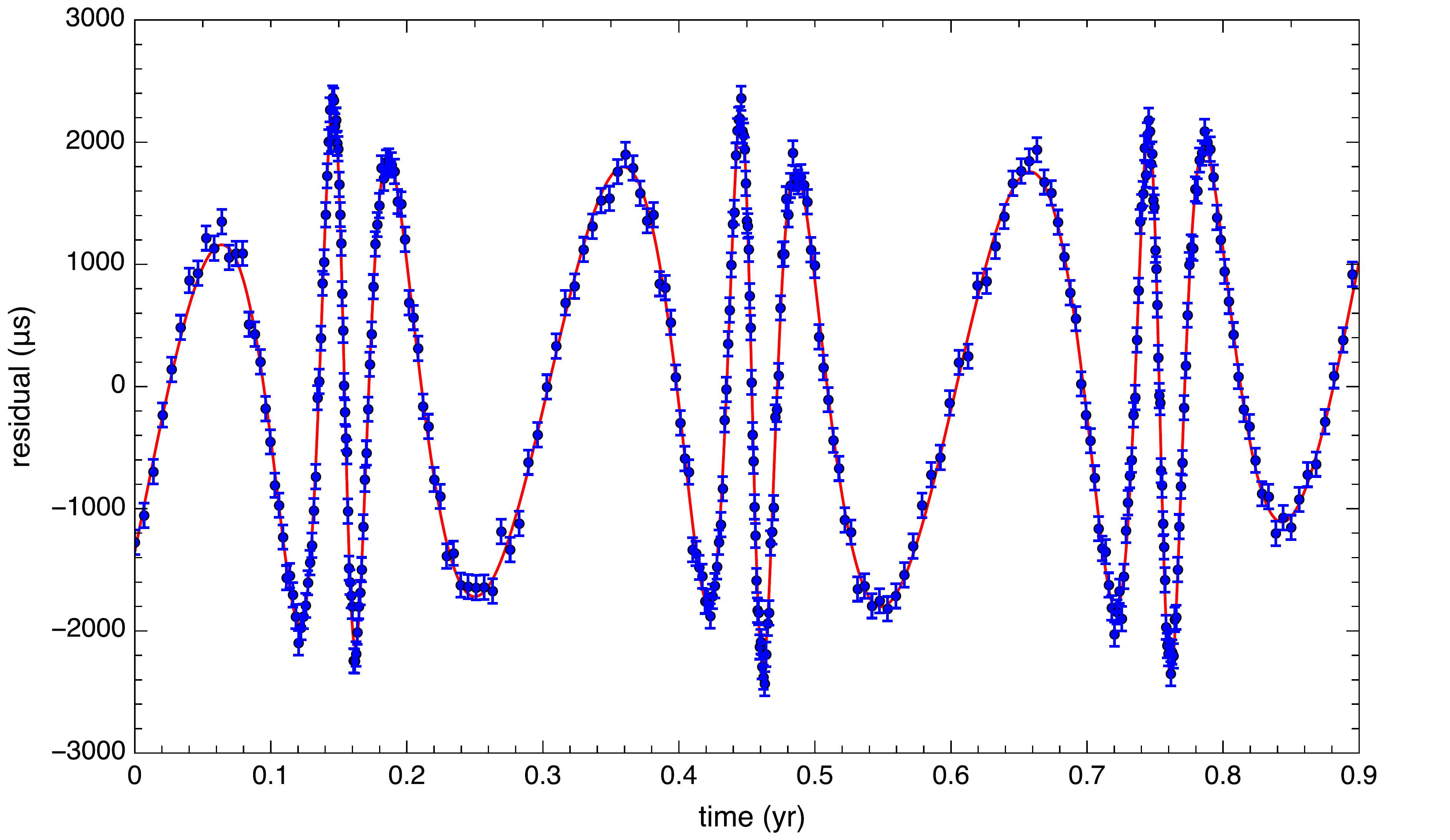}
  \caption{Characteristic pulsar timing residuals caused by the
    quadrupole moment of Sgr A* BH from simulated TOAs for three
    orbital phases \citep{lwk+12} (figure courtesy of N. Wex).  The
    orbital period is assumed to be $P_b=0.3$\,yr and the eccentricity
    $e=0.5$.  The red line is the fitted model.  \label{fig:q_res}}
\end{figure}

BHs are predicted to be the ultimate outcome of massive stars in the
theory of GR.  Evidence for the existence of BHs in various
astrophysical environments has accumulated during the past few
decades, and is now rather robust \citep{nm13}.  However, such
observations are still indirect, and no high-precision tests of BHs
are yet possible.  Testing BH physics is a fantastic and intriguing
goal in the fields of astronomy and physics. The goal of precision
tests of properties of BHs will be achieved with the SKA
\citep{kbc+04}.

The combination of the sensitivity and the large field of view of the
SKA will inevitably multiply the number of pulsars by tens to
hundreds, making detection of rare objects like PSR-BH binaries
possible.  The shorter integration time in pulsar search (say, a
10-min integration as proposed) will be beneficial to discoveries of
extremely relativistic binaries.  A relativistic PSR-BH system in a
clean environment represents a ``holy grail'' of pulsar astronomy.  A
vast unexplored parameter space in terms of gravitational potential,
spacetime curvature and object compactness will then become accessible
(see Figure~\ref{fig:grav}).  Pulsar -- stellar-mass BH binaries,
pulsar -- intermediate-mass BH binaries, and pulsars orbiting the Sgr
A* BH probe different new regimes of gravity where no other gravity
experiments are yet able to be performed with precision tests
\citep{wk99,pl04}.  The pulsar -- stellar-mass BH binary may naturally
come from a primary binary composed of a $>20\,M_\odot$ massive star
and another $>8\,M_\odot$ star, where the former ends up as a BH and
the latter ends up as a NS.  The pulsar -- intermediate-mass BH
binaries may exist in the centre of some globular clusters
\citep{csc14,hpb+14}. Given the dense environment in the Galactic
centre, there may be numerous NSs orbiting the Sgr A* BH, some of
which may be detected as radio pulsars \citep{pl04}.  However, no
pulsars close enough to Sgr A* BH for gravity tests have been detected
yet.  Recently, a magnetar was detected through X-ray
\citep{mgz+13,kbk+13} and radio \citep{efk+13} observations, however,
it is not close enough to the Sgr A* BH for precision tests of gravity
\citep{efk+13,elc+14}.  Because of scattering by the interstellar
medium, in general we will need the SKA2 to detect these pulsars in
the Galactic centre \citep{wcc+12,wle+13,elc+14}, although this will
depend on which bands are included in SKA1-MID, for example, the
inclusion of Band 5 will greatly enhance the prospects for doing
this. For detecting the other two types of PSR-BH binaries, the SKA1
will start to contribute significantly.

In the case of relativistic PSR-BH binaries, the prominent effects
from gravitation will, on one hand, reinforce better measurements of
existing tests such as the Shapiro delay and the gravitational wave
damping, and on the other hand, enable completely new tests of the
``cosmic censorship conjecture'' and the ``no-hair theorem''
\citep{de98,wk99,lwk+12,lewk14}.  The ``cosmic censorship conjecture''
states that for a realistic BH, there always exists an event horizon
which prevents us from looking into its central singularity, thus
preserving classical predictability in a spacetime. It imposes a
maximum spin, $S_{\rm max} = GM^2/c$, on a rotating BH with a given
mass $M$. The dimensionless spin of a BH, $\chi \equiv c S / GM^2$,
should always satisfy the constraint $\chi \leq 1$ \citep{mtw73}. In
alternative gravitational theories, this bound may be slightly
violated, as in the Einstein-Gauss-Bonnet theory, where, besides the
Einstein-Hilbert term, the Gauss-Bonnet topological term is included
in the Lagrangian of gravity \citep{kkr11}.  The ``no-hair theorem''
states the uniqueness of a Kerr BH after it settles down. It relates
the higher-order moments of a BH exclusively to its mass and spin. For
example, if the ``no-hair theorem'' is valid, the dimensionless
quadrupole of a BH, $q \equiv c^4 Q / G^2M^3$ where $Q$ is the
quadrupole moment of BH, should satisfy $q = -\chi^2$ \citep{mtw73}.
If a complex, massive scalar field, minimally coupled to GR, is
introduced, then the ``no-hair theorem'' could be violated
\citep{hr14}.

For a relativistic PSR -- Sgr A* BH system, the pulsar can be treated
as a test particle in the first-order approximation \citep{hl08,hl12}.
The strategy of performing the tests of the ``cosmic censorship
conjecture'' and the ``no-hair theorem'' was investigated in
\citet{lwk+12}.  First of all, as usual in pulsar timing, we can
estimate the mass of the Sgr A* BH through the measurement of the
periastron advance rate.  After that, a better mass determination and
the geometry information can be obtained with the detection of the
Shapiro delay and the Einstein delay. With the help of the
frame-dragging effects (see section~\ref{sec:binary}), the spin-orbit
coupling then allows an initial extraction of an upper limit of the
spin magnitude of the Sgr A* BH. Thus an initial test of the ``cosmic
censorship conjecture'' becomes available by looking at the magnitude
of $\chi$.  This may enable us to look into the bizarre possibility of
Sgr A* BH being a boson star \citep{kks12}.  With a better
determination of the Sgr A* BH mass, the separation of the impact on
the periastron advance from the BH spin is made plausible.  This
further determines the 3-dimensional spin of the BH with magnitude and
direction.  An extraction of the periodic effects of the quadrupole
from timing residuals will then allow a quantitative test of the
``no-hair theorem''.  The timing residual from the quadrupole of Sgr
A* BH has a characteristic shape, as illustrated in
Figure~\ref{fig:q_res} for a hypothetical system with an orbital
period $P_b \simeq 0.3$\,yr and an eccentricity $e\simeq0.5$
\citep{lwk+12}. According to simulations for an orbit with a period of
several months, the spin and the quadrupole can be measured with
precisions of $\sim0.1\%$ and $\sim1\%$ respectively after five years
of observations with the SKA. Therefore, the test of the ``no-hair
theorem'' can be performed at the $\sim1\%$ level by then
\citep{lwk+12}.

Similarly, in the cases of stellar-mass BH and intermediate-mass BH
binaries, the spin information of the BH can be extracted with the
help of frame dragging as well if the orbit is tight enough
\citep{wk99}.  For the stellar-mass BH companion, a relativistic
binary can provide much better constraints on the scalar-tensor
theories than the tests with gravitational waves from inspiralling
binaries \citep{de98}. These constraints will also be an order of
magnitude better than the expected near-future constraint from the
GAIA satellite when $\beta_0$ is negative or small; see
Figure~\ref{fig:scalartensor} for results from a simulated PSR-BH
system \citep{lewk14}.

\section{Conclusion and Outlook}

By the 2020s, radio astronomy will undergo a revolution brought by the
onset of the SKA1. Its pulsar survey will deliver a large population
of pulsars, among which there will be $\sim1800$ MSPs. The most stable
ones of them will be used to perform a number of precision gravity
tests. The improvement in the sensitivity of the timing precision
allows us a closer look at many aspects of the foundation of gravity.
In the era of the SKA, the existing tests will be pushed to new
extremes, and completely new tests will also be made possible with
systems including NS-NS binaries, NS-WD binaries and NS-BH binaries.
Advances from the theoretical side will also contribute to the area of
experimental gravity with novel tests. It is indeed a virtue of nature
that pulsar binaries of different types are equipped with different
characteristics to perform different tests with.  For example, the
NS-NS systems provide extremely precise tests of GR, in particular
periodic, non-dissipative effects in orbital dynamics ({\it e.g.} with
the Einstein delay and general-relativistic orbital deformation),
while the NS-WD and NS-BH systems are suitable to study the
dissipative radiative property of gravity in the presence of asymmetry
in gravitational binding energy ({\it e.g.} tests of the gravitational
dipole radiation and the constancy of the gravitational
constant). Furthermore, PSR-BH binaries will be useful to test the BH
physics ({\it e.g.} the ``cosmic censorship conjecture'' and the
``no-hair'' theorem).  Tests of gravitational theories from many
facets in the era of the SKA will profoundly advance our
understandings of fundamental physics.

The output of gravity tests with pulsars has two possibilities.  In
the first possibility, GR is falsified. This will certainly provide an
important hint for our further understanding of the fundamental
interaction, gravity, and the nature of spacetime. In the second
possibility, GR is confirmed to new precision with flying
colours. This will further reinforce our confidence in using GR as a
tool to investigate other phenomena, for example, using GR effects to
make mass measurements of binary pulsars, helping to deepen our
understandings of non-perturbative QCD effects \citep{wxe+14}. It also
casts experimental constraints on constructing new gravitational
theories in relevant areas, for example, to explain the nature of dark
matter and dark energy. In both cases, tests of gravitational theories
with great precision will provide important clues to reconcile GR and
quantum principles.

As a final remark, with the better timing capability and new
discoveries of stable pulsars from the SKA, the pulsar timing array
(PTA) projects will leap towards a new era as well.  Complementary to
the strong-field tests discussed here, PTAs make tests of
gravitational theories with gravitational waves in the radiative
regime available \citep{jhm+14}.

\section*{Acknowledgement}

We thank Norbert Wex for kindly providing figures and carefully
reading the manuscript.


\end{document}